\documentclass[prd,twocolumn,preprintnumbers,showpacs,nofootinbib,superscriptaddress,notitlepage,floatfix]{revtex4-1}
\usepackage{hyperref}
\usepackage{graphicx}
\usepackage{amsmath}
\usepackage{amssymb}
\usepackage{slashed}
\usepackage{amsfonts}
\usepackage{xcolor}
\usepackage{multirow}
\usepackage[caption=false]{subfig}
\usepackage{array}
\usepackage{mwe}
\usepackage{mathrsfs}

\usepackage[utf8]{inputenc}
\usepackage[normalem]{ulem}

\DeclareRobustCommand{\eq}[1]{Eq.~\eqref{eq:#1}}

\DeclareRobustCommand{\eqs}[2]{Eqs.~\eqref{eq:#1} and \eqref{eq:#2}}

\DeclareRobustCommand{\fig}[1]{Fig.~\ref{fig:#1}}

\DeclareRobustCommand{\eq}[1]{Eq.~(\ref{eq:#1})}
\DeclareRobustCommand{\eqs}[2]{Eqs.~(\ref{eq:#1}) and (\ref{eq:#2})}

\newcommand{\nn}{\nonumber}

\newcommand\bets{\begin{table*}}
\newcommand\eets[1]{\label{tb:#1}\end{table*}}

\begin{document}

\title{Comments on ``Non-local Nucleon Matrix Elements in the Rest Frame''}

\author{Xiang Gao}
\affiliation{Physics Division, Argonne National Laboratory, Lemont, IL 60439, USA}

\author{Jinchen He}
\affiliation{Department of Physics, University of Maryland, College Park, MD 20742, USA}
\affiliation{Physics Division, Argonne National Laboratory, Lemont, IL 60439, USA}

\author{Yushan Su}
\affiliation{Department of Physics, University of Maryland, College Park, MD 20742, USA}

\author{Rui Zhang}
\affiliation{Physics Division, Argonne National Laboratory, Lemont, IL 60439, USA}

\author{Yong Zhao}
\affiliation{Physics Division, Argonne National Laboratory, Lemont, IL 60439, USA}

\begin{abstract}
In a recent paper, ``\textit{Non-local Nucleon Matrix Elements in the Rest Frame}''~\cite{Karpie:2024bof}, it was observed that the next-to-leading order calculations of the renormalization factor can describe, to a few percent accuracy, the logarithm of the lattice QCD rest frame matrix elements with separations up to distances of 0.6 fm on multiple lattice spacings. We argue that perturbative QCD breaks down at such a distance scale after resumming the associated large logarithms, while the \textit{ansatz} used in the analysis there is not justified in perturbation theory. Besides, we explain the observation in Ref.~\cite{Karpie:2024bof} and demonstrate that the \textit{ansatz} fails to describe the data for $z>0.3$ fm, showing an opposite trend. Finally, although Ref.~\cite{Karpie:2024bof} proposes multiplying the \textit{ansatz} by a Gaussian correction model, which is shown to reduce the discrepancy with the data, this does not legitimize the use of perturbative QCD at such distance scales.
\end{abstract}

\maketitle

In a recent paper, ``\textit{Non-local Nucleon Matrix Elements in the Rest Frame}''~\cite{Karpie:2024bof}, the authors studied the static nucleon matrix elements of the gauge-invariant quark bilinear operator
\begin{align}
    O_{\gamma^t}(z) = \bar{\psi}(z) \gamma^t W(z,0) \psi(0)\,,
\end{align}
where $W(z,0)$ is a spatial Wilson line, from lattice QCD. The matrix elements were computed at an unphysical pion mass $m_\pi\sim 445$ MeV with three different lattice spacings. 
They were compared to the \textit{ansatz}
\begin{align}\label{eq:ansatz}
\exp\left(\Gamma({z/a_{\rm PR}})\right)\,,
\end{align} 
where $\Gamma$ is the next-to-leading order (NLO) ultraviolet (UV) renormalization factor~\cite{Radyushkin:2017lvu} under the Polyakov regularization (PR)~\cite{Polyakov:1980ca},
\begin{align}\label{eq:nlo}
    \Gamma(\zeta) &= {\alpha_s C_F\over 2\pi}\bigg[\Big(2 + {1\over \zeta^2}\Big) \ln (1+\zeta^2) -{1\over2}\ln\zeta^2 \nn\\
    &\qquad  - 2\zeta \tan^{-1}\zeta - 1 \bigg]\,,
\end{align}
with $C_F=4/3$, and the strong coupling constant $\alpha_s$ is a fitting parameter. The coupling $\alpha_s$ depends on the PR regulator, $1/a_{\rm PR}$, which was identified as the lattice UV cutoff $\pi/a$ that regulates both the linear and logarithmic divergences~\cite{Chen:2016fxx,Karpie:2024bof}.
The $\alpha_s$ is given by the leading-order (LO) approximation~\cite{Karpie:2024bof}
\begin{align}
	\alpha_s= \alpha_s({\pi/ a}) = - {2\pi \over (11-n_f/3) \ln( a\Lambda_{\rm QCD}/\pi)}\,,
\end{align}
where $n_f$ is the number of active quark flavors, and $\Lambda_{\rm QCD}$ becomes the fitting parameter. The lattice spacing $a$ should be much smaller than $1/\Lambda_{\rm QCD}$ for perturbation theory to be valid. Ref.~\cite{Karpie:2024bof} also studied an \textit{ansatz} based on lattice perturbation theory, which led to similar findings, so we only focus on the \textit{ansatz} in \eq{ansatz} for discussion here.

By neglecting the correlation of matrix elements at different $z$ on the same lattice ensemble, the authors found that the simple \textit{ansatz} in \eq{ansatz} with $\Lambda_{\rm QCD}\approx 250$ MeV appears to pass through almost all the data points, despite that the $\chi^2$/dof is of $O(10)$ or larger due to high data precision. 
They concluded that when the lowest two separations are neglected, both of the perturbative results agree with the data for the logarithm of the matrix element to within 5\%.

While it is nontrivial to observe such agreement, it does not mean that perturabtive QCD can still be used at such a large distance scale.
Above all, there is a plethora of evidence showing that perturbation theory breaks down beyond a distance scale of $0.2\text{–}0.3$ fm, such as QCD sum rules~\cite{Shifman:1978bx}, the instanton model~\cite{Schafer:1996wv}, gluon effective mass~\cite{Gongyo:2013sha,Schrock:2015pna}, the static potential~\cite{Pineda:2002se}, etc. For further discussions see Refs.~\cite{Ji:2020brr,Ji:2020byp,Ji:2022ezo}. Therefore, the hadronic matrix elements of nonlocal operators like $O_{\gamma^t}(z)$ should not be exceptional.

Now let us turn to the \textit{ansatz} in \eqs{ansatz}{nlo}. There are several issues regarding its validity:

\begin{enumerate}

\item The naive exponentiation of the NLO correction is not justified. Except for the linear divergence~\cite{Ji:2017oey,Ishikawa:2017faj,Green:2017xeu}, exponentiation is usually the outcome of solving the renormalization group equation (RGE), which is known as resummation. The logarithmic terms can be resummed as they follow the RGE, but the finite terms are either not resummable or follow completely different RGEs. Besides, in QCD no resummation can be done without the running of $\alpha_s$. The \textit{ansatz} in \eq{ansatz} does not meet any of these conditions, so it is not a perturbative solution. Note that Ref.~\cite{Karpie:2024bof} compared the logarithm of the matrix element to the \textit{ansatz}, which, however, is equivalent to exponentiating the latter.

\item The resummation requires running $\alpha_s$ from an initial scale which is proportional to $1/z$~\footnote{The common choice of the initial scale in the literature~\cite{Peskin:1995ev,Schafer:1996wv,Pineda:2000gza,Pineda:2002se} is $c/z$ with $c\approx1$, which is varied within a range to estimate the associated uncertainty. One may argue that if $c\approx 3$, then perturbation theory may converge at $z\sim 0.6$ fm, but there is no evidence to support that $c$ is substantially different from one and, moreover, independent of the external state.} for the nonlocal matrix element~\cite{Gao:2021hxl}. When $\alpha_s (1/z) \sim 1$, or $z\sim 0.3$ fm, perturbation theory already breaks down, let alone converging, which has been demonstrated in Refs.~\cite{Gao:2022iex,Su:2022fiu}. 

\item The NLO correction did not account for the leading renormalon when regulating the linear divergence~\cite{Beneke:1998ui,Bauer:2011ws}, so perturbative convergence is not guaranteed. It has been demonstrated that without subtracting this renormalon, the $\overline{\rm MS}$ version of $\Gamma(\zeta)$ already fails to converge at next-to-next-to-leading order at $z= 0.2\text{–}0.3$ fm~\cite{Gao:2021dbh,Zhang:2023bxs}.

\end{enumerate}

Without resummation, the perturbative expression in \eq{nlo} can always be evaluated at any $z$, but it does not carry much meaning in the non-perturbative region. Besides, since all the other effects, including higher order corrections, discretization, non-perturbative contamination, etc., are absorbed into the single parameter $\alpha_s$ or $\Lambda_{\rm QCD}$, it loses universality and predictive power. Therefore, the \textit{ansatz} at large $z$ is essentially a model whose agreement with data is neither sufficient nor necessary to justify the use of perturbation theory.

Then the question is: why does a simple one-parameter model seem to agree with the lattice data well?

First of all, we should be aware that the metric for ``agreement'' in Ref.~\cite{Karpie:2024bof} is not the usual correlated $\chi^2$/dof, but the relative difference between the data and fit, which was found to be at percent level or even less. One might argue that this is not sufficient evidence for an agreement, but the small difference still suggests that the model must have captured the key physics.

\begin{figure}
    \centering
    \includegraphics[width=0.8\linewidth]{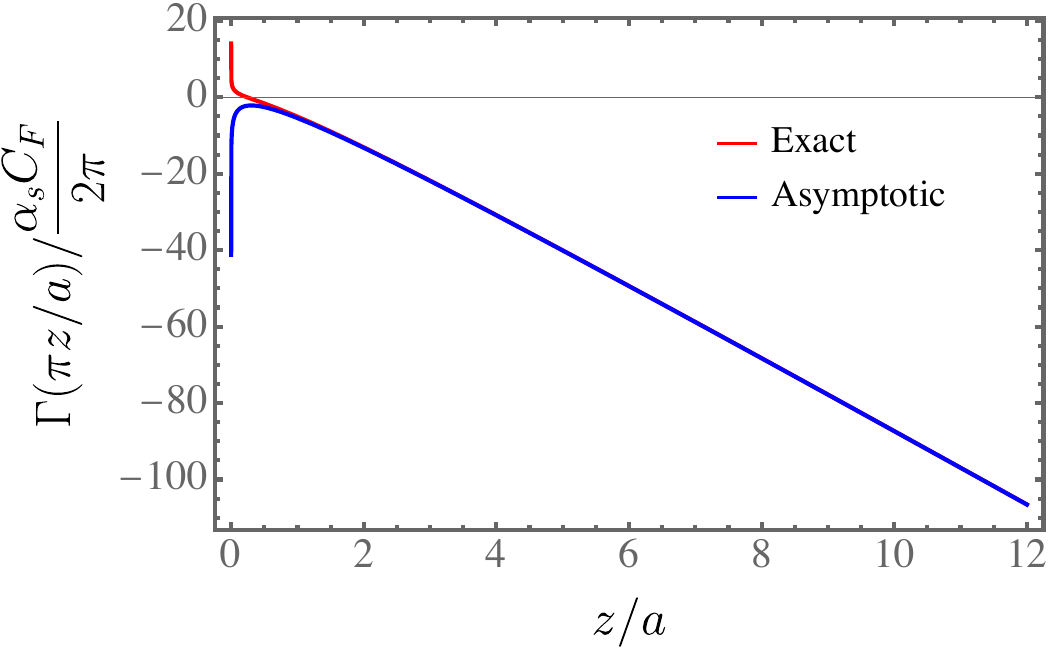}
    \caption{The exact and asymptotic solutions to $\Gamma(\pi z/a)$. 
    }
    \label{fig:gz}
\end{figure}

To answer this question, let us take a closer look at the \textit{ansatz}. The Polyakov regulator $a_{\rm PR}= a/\pi$ can be regarded as a reduced lattice spacing, so the limit $\zeta=z/ a_{\rm PR}\gg1$ is quickly saturated as $z$ increases,
\begin{align}\label{eq:asym}
	\lim_{\zeta\to\infty} \Gamma(\zeta) &= {\alpha_s C_F\over 2\pi}\left( - \pi |\zeta| + {3\over2}\ln\zeta^2 + 1 \right)\,,
\end{align}
where the linear term inside the brackets is exactly the power divergence. \fig{gz} compares the exact and asymptotic solutions to $\Gamma(\pi z/a)$, which converge fast for $z\ge 2a$, or, almost all the analyzed data points.

The \textit{ansatz} in \eq{asym} is similar to that used in the self-renormalization scheme~\cite{LatticePartonLPC:2021gpi},
\begin{align} \label{eq:sr}
	\Gamma_{\rm SR}(z,a) &= {kz \over a \ln ( a\Lambda_{\rm QCD})} + m_0 z + \ln[Z_{\overline{\rm MS}}(z)] + \ldots\,,
\end{align}
where the first term is the linear divergence, and the parameters $k$ and $\Lambda_{\rm QCD}$ are properly chosen or fitted to effectively take into account part of the higher order corrections to it~\cite{Lepage:1992xa}. The short-distance coefficient $Z_{\overline{\rm MS}}(z)$ is in the $\overline{\rm MS}$ scheme. Here $m_0$ is the leading renormalon and of $O(\Lambda_{\rm QCD})$, which was not considered in Ref.~\cite{Karpie:2024bof}, and $\ldots$ stands for discretization effects and the overall $z$-independent renormalization factor, which are unimportant for our discussion here. According to the renormalization relation of the nonlocal operator~\cite{Ji:2017oey,Ishikawa:2017faj,Green:2017xeu}, the linear divergence depends only on the UV regulator and can be exponentiated. More importantly, this relation holds for all $z$. In contrast, $Z_{\overline{\rm MS}}(z)$ characterizes the short-distance behavior of the matrix elements and becomes invalid for $z>0.3$ fm.

In Ref.~\cite{LatticePartonLPC:2021gpi}, the static pion matrix elements of $O_{\gamma^t}(z)$ was computed with five MILC ensembles~\cite{MILC:2012znn} and three RBC/UKQCD ensembles~\cite{RBC:2014ntl} at different lattice spacings. The linear divergence in \eq{sr} can be extracted through the $a$-dependence of the matrix elements without constraining $m_0$. However, it was found that $k$ and $\Lambda_{\rm QCD}$ are strongly correlated. Different values of the duo can achieve a ``good'' fit while shifting the coefficient of $z$ by constants, which cannot be distinguished from $m_0$ as it is the nature of renormalon ambiguity. Therefore, the \textit{ansatz} in \eq{asym} is equivalent to fixing $k$, and there must be a value of $\Lambda_{\rm QCD}$ that can fit the data.
Then, $m_0$ is fitted from the short-distance matrix elements at $a\ll z \lesssim 0.2$ fm for a chosen pair of $(k,\Lambda_{\rm QCD})$. However, after subtracting the linear divergence and $m_0z$, the perturbative prediction $Z_{\overline{\rm MS}}(z)$ fails to describe the data beyond $z=0.2\text{–}0.3$ fm for any choice of $(k,\Lambda_{\rm QCD})$~\cite{LatticePartonLPC:2021gpi}. One should also expect \eq{asym} to miss the data points at long distances. Note that \eq{sr} has the same issue of lacking resummation, higher-order correction, and the regulation of leading renormalon, but even if they are all included~\cite{Holligan:2023rex,Zhang:2023bxs}, it will be more obvious to see perturbation theory break down at $z\gtrsim 0.2\text{–}0.3$ fm.

\begin{figure}
    \centering
    \includegraphics[width=0.8\linewidth]{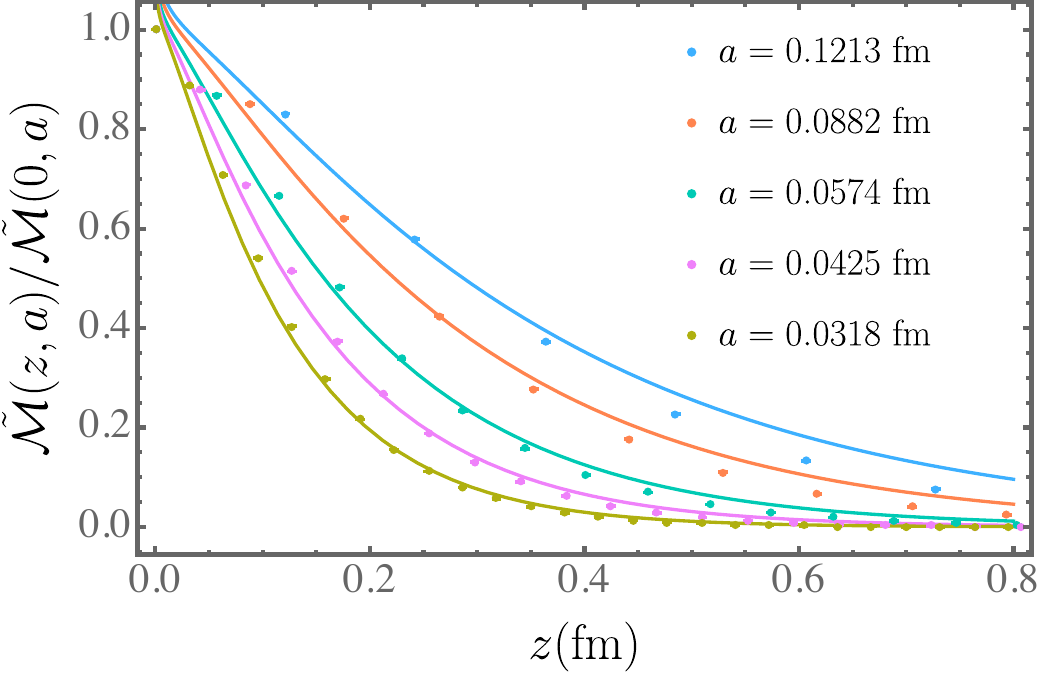}
    \includegraphics[width=0.8\linewidth]{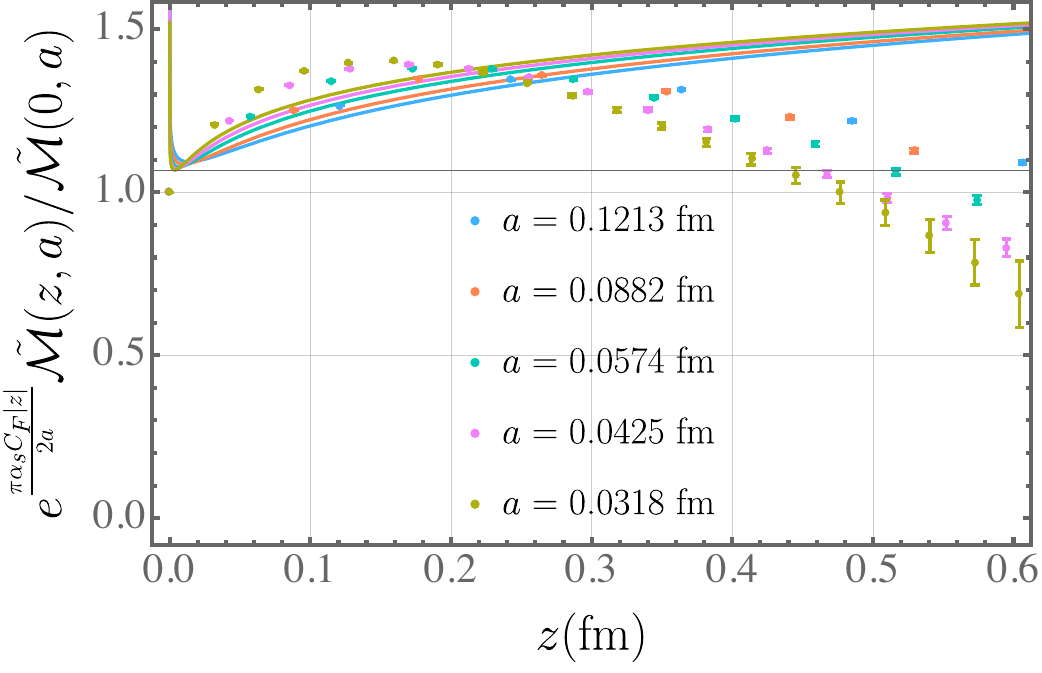}
    \caption{The static pion matrix elements (data points), computed in Ref.~\cite{LatticePartonLPC:2021gpi} from the MILC ensembles, are compared to the fits (curves) with the \textit{ansatz} in \eq{ansatz}, which corresponds to $n_f=4$ and $\Lambda_{\rm QCD}=0.12$ GeV. Upper panel: the full matrix elements. Lower panel: the matrix elements after subtracting $\exp[- \alpha_s C_F/(2\pi) \cdot \pi^2 |z|/a ]$.
    }
    \label{fig:comp}
\end{figure}

\begin{figure}
    \centering
    \includegraphics[width=0.8\linewidth]{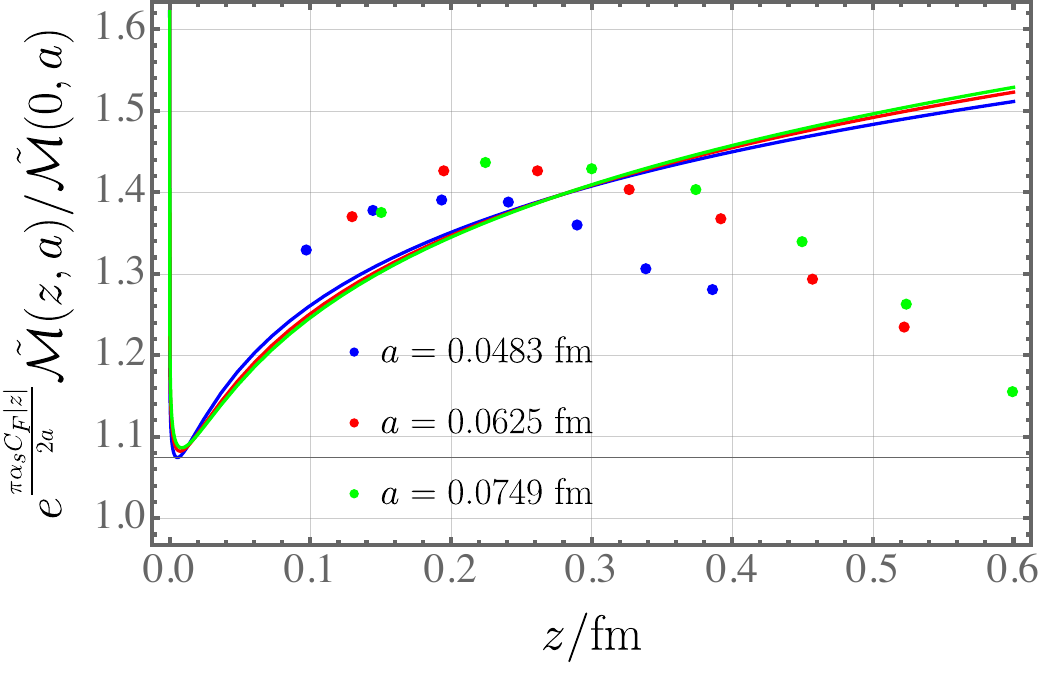}
    \caption{After subtracting the linear divergence, the static proton matrix elements (data points) in Ref.~\cite{Karpie:2024bof} are compared to the theory (curves) with $n_f=2$ and $\Lambda_{\rm QCD}$ fitted from each ensemble with $z\ge 2a$. Only the central values are plotted.
    }
    \label{fig:nucleon}
\end{figure}

To verify our claims, we subtract the linear term in \eq{asym} from both the lattice matrix elements $\tilde {\cal M}(z,a)$ and $\exp\left[\Gamma(\zeta)\right]$, and then compare them. As shown in \fig{comp}, the fits with $n_f=4$ and $\Lambda_{\rm QCD}=0.12$ GeV pass through most of the MILC data points from Ref.~\cite{LatticePartonLPC:2021gpi} up to fairly long distances~\footnote{We found that to obtain a reasonable $\Lambda_{\rm QCD}$, one must use unsmeared Wilson lines to generate the matrix elements, otherwise the fitted $\Lambda_{\rm QCD}$ would be incredibly small ($\ll 10^{-3}$GeV). This is expected as smearing reduces the UV fluctuations, including the linear divergence.}. But the \textit{ansatz} fails to agree with data with an opposite trend at $z\gtrsim 0.3$ fm, which is consistent with the findings of Ref.~\cite{LatticePartonLPC:2021gpi}. Even if we shift the value of $\Lambda_{\rm QCD}$ by 30\% to account for the fitting error, the disagreement still persists at similar distance scales.
Figure.~\ref{fig:nucleon} compares the nucleon matrix elements in Ref.~\cite{Karpie:2024bof} to the fits with $n_f=2$ and $\Lambda_{\rm QCD}\approx 0.25$ GeV~\footnote{The data are obtained by reading off Fig.~5 of Ref.~\cite{Karpie:2024bof}.}. They also diverge at $z\gtrsim 0.3$ fm with opposite trends, showing that the \textit{ansatz} fails to describe the data. Notably, the relative error between the data and the theory can be 15\%, in contrast to the 5\% reported in Ref.~\cite{Karpie:2024bof}. This is because the latter plotted the ratio between the logarithms of the data and the theory instead, which suppressed their discrepancy, but statistically there is no difference.

Therefore, the linear divergence dominates the bare static matrix element at large distances. Since the linear divergence only depends on the UV cutoff and is universal for all $z$, it is unsurprising that a simple parametrization based on the NLO expression can approximate it well. However, the remaining contribution to the matrix element varies mildly with $z$, and its deviation from the \textit{ansatz} is masked by the dominance of the linear divergence. This explains the apparent agreement between the data and theory observed in Ref.~\cite{Karpie:2024bof}.

Finally, a Gaussian correction model was multiplied with \eq{ansatz} in Ref.~\cite{Karpie:2024bof}, which further reduces the discrepancy and even brings the $\chi^2$/dof below one for one of the data sets. This was taken as support for the hypothesis in the pseudo-PDF approach~\cite{Radyushkin:2017lvu,Orginos:2017kos} that the zero-momentum matrix element can be factorized into a product of a perturbative part and a non-perturbative Gaussian correction. It may also be used to justify the use of perturbation theory for ratio-scheme renormalized matrix elements~\cite{Orginos:2017kos}, where the Gaussian correction cancels, at $z \sim 0.6$ fm.

However, since the breakdown of perturbative QCD at such distance scales is well established and independent of lattice calculations, and given that exponentiating the NLO correction is also not justified, it is incorrect to interpret the model as perturbation theory with minor corrections. Instead, the perturbative \textit{ansatz}, including both the coupling $\alpha_s$ and its functional form, should be regarded as part of the modeling itself. Because the residual $z$-dependence is mild once the linear divergence is properly parametrized, and the data uncertainty increases with $z$, it is unsurprising that adding more parameters to the model can improve the fit quality.

\vspace{0.3cm}
In summary, the \textit{ansatz} in Ref.~\cite{Karpie:2024bof} is not a valid perturbative solution beyond $z = 0.2\text{–}0.3$ fm, a fact that is well known and independent of lattice calculations. Moreover, it cannot be made legitimate by multiplying a correction model, despite that the latter may reduce the discrepancy with the data.
The similarity observed between the fit and data is mainly due to the dominance of the linear divergence, which was adequately parametrized by the linear term in the \textit{ansatz}. But apart from the linear divergence, the \textit{ansatz} cannot describe the data beyond $\sim 0.3$ fm, which is consistent with knowledge from the literature.
Finally, we emphasize that a rigorous comparison of perturbation theory and lattice matrix elements must be done with resummation and the regulation of leading renormalon~\cite{Pineda:2002se,Holligan:2023rex,Zhang:2023bxs}, which, after all, is still valid at short distances only.

\begin{acknowledgments}

We thank the Lattice Parton Collaboration for providing the static pion matrix elements.
This material is based upon work supported by the U.S. Department of Energy, Office of Science, Office of Nuclear Physics through Contract No.~DE-AC02-06CH11357, and the \textit{Quark-Gluon Tomography (QGT) Topical Collaboration} with Award No.~DE-SC0023646.

\end{acknowledgments}

\bibliography{Refs}

\end{document}